# Dirac semimetal PdTe$_2$ temperature-dependent quasiparticle dynamics and electron-phonon coupling

Shu-Yu Liu,[1] Shuang-Xing Zhu,[1] Qi-Yi Wu,[1] Chen Zhang,[1] Peng-Bo Song,[2, 3] You-Guo Shi,[2, 3]
Hao Liu,[1] Zi-Teng Liu,[1] Jiao-Jiao Song,[1] Fan-Ying Wu,[1] Yin-Zou Zhao,[1] Xiao-Fang Tang,[1]
Ya-Hua Yuan,[1] Han Huang,[1] Jun He,[1] H. Y. Liu,[4] Yu-Xia Duan,[1] and Jian-Qiao Meng[1, 5, *]

[1] *School of Physics and Electronics, Central South University, Changsha 410083, Hunan, China*
[2] *Beijing National Laboratory for Condensed Matter Physics,*
*Institute of Physics, Chinese Academy of Sciences, Beijing 100190, China*
[3] *School of Physical Sciences, University of Chinese Academy of Sciences, Beijing 100049, China*
[4] *Beijing Academy of Quantum Information Sciences, Beijing 100085, China*
[5] *Synergetic Innovation Center for Quantum Effects and Applications (SICQEA),*
*Hunan Normal University, Changsha 410081, China*
(Dated: Thursday 2$^{nd}$ September, 2021)

Dirac semimetal PdTe$_2$ single-crystal temperature-dependent ultrafast carrier and phonon dynamics were studied using ultrafast optical pump-probe spectroscopy. Quantitative analysis revealed a fast relaxation process ($\tau_f$) occurring at a subpicosecond time scale originating from electron-phonon thermalization. This rapid relaxation was followed by a slower relaxation process ($\tau_s$) on a time scale of $\sim$ 7-9.5 ps which originated from phonon-assisted electron-hole recombination. Two significant vibrational modes resolved at all measured temperatures. These modes corresponded to in-plane ($E_g$), and out-of-plane ($A_{1g}$), Te atoms motion. Test results suggested that pure dephasing played an important role in the relaxation processes. Analysis of the electron-phonon coupling constant suggested that the $A_{1g}$ mode contributes greatly to the superconductivity, and high-frequency phonons are also involved forming of Cooper pairs. Our observations should improve the understanding of complex superconductivity of PdTe$_2$.

## I. INTRODUCTION

Recently, layered, transition-metal dichalcogenides (TMDCs) have been extensively studied for their electronic and photoelectric qualities in such fields as superconductivity, nontrivial topological properties, charge density waves, extremely large positive magnetoresistance, and others [1–5]. Topological superconductors attracted extensive research interest [6, 7], after the discovery of topological insulators [8, 9]. Tuning the topological superconductivity states, such as by tuning Majorana fermions, was thought might lead to a new quantum computation class. Several different physical properties coexist simultaneously in TMDCs [10, 11]. Theoretical calculations and experimental results suggest that several TMDCs may possess topological superconductivity [12–14].

PdTe$_2$ is a favored research component due to its valuable physical properties. Its superconductivity, with transition temperatures of $T_c \sim$ 1.7 K, was discovered in the early 1960s [15]. Recently, angle-resolved photoemission spectroscopy (ARPES) confirmed that PdTe$_2$ is either a type-II, or a type-I and type-II co-existent, Dirac semimetal with a spin-polarized topological surface state [16–21]. This topological surface state commends PdTe$_2$ for consideration as possible topological superconductor [22, 23]. However, heat capacity, penetration depth, and other tests suggest that PdTe$_2$ is a fully-gapped, $s$-wave conventional superconductor [24, 25]. Superconductors may be classified as either type-I and type-II superconductor depending upon the presence, or absence, of an intermediate phase of mixed ordinary and superconducting properties. Some studies suggest that bulk PdTe$_2$ is a typical type-I superconductor ($T_c \sim$ 1.65 K ) [23, 26, 27]. Its surface has a mix of type-I and type-II superconductivity as surface superconducting transition temperature of $T_c^s =$ 1.33 K [23, 24, 27, 28]. Intrinsic surface electronic inhomogeneities was thought to be responsible for the mixed type I and type II superconducting behaviors. For conventional superconductors, Cooper pairs are bonded by phonons. Currently neutron inelastic scattering [29] and theoretical calculations [30] suggest that phonon dispersion near $\Gamma$ is important for PdTe$_2$ superconductivity. Investigating PdTe$_2$ phononic properties holds promise for the experimental determination of the electron-phonon ($e$-$ph$) coupling constant $\lambda$.

Ultrafast optical pump-probe spectroscopy is a powerful tool for studying quasiparticle dynamics especially temperature evolution behavior of low energy scale collective modes [31–35]. ARPES is a powerful tool for detecting the many-body effects which have yet to identify PdTe$_2$ phonons [16–21]. However, detecting, observing, and measuring this temperature evolution behavior is usually far beyond ARPES energy resolution limits.

Ultrafast optical pump-probe spectroscopy measurements were used to investigate the $e$-$ph$ coupling constant $\lambda$ of PdTe$_2$ crystals in order to gain insight into its super-conductivity. Two distinct carrier relaxation processes with different time scales were observed. They originated from $e$-$ph$ thermalization ($\tau_f$) and phonon-assisted electron-hole ($e$-$h$) recombination ($\tau_s$), respectively. Two optical phonons, in-plane ($E_g$), and out-of-plane ($A_{1g}$) Te atoms motion, were detected at temperatures ranging from 5 - 300 K. $E_g$ energy is consistent with that of a $\Gamma$ phonon which is thought to be responsible for PdTe$_2$ superconductivity [29, 30]. $e$-$ph$ coupling constant analysis reveals that high frequency phonons are involved in the Cooper pair formation.

## II. EXPERIMENT

The flux method was used to grow high quality PdTe$_2$ single crystals. High purity Pd and Te with a molar ratio of 1:4 were placed in an alumina crucible and sealed in a quartz tube. The tube was heated to 800 °C and then slowly cooled to 500 °C. Excess Te was removed by centrifuge. PdTe$_2$ single crystal remained at the bottom of the crucible. Crystals as large as $\sim$ 4 × 4 mm$^2$ are obtained.

Ultrafast time-resolved differential reflectivity $\Delta R/R$ was performed on a high-quality PdTe$_2$ single crystal with a pulse laser produced by a Yb-based femtosecond (fs) laser oscillator. Pulse width was $\sim$35 fs with a center wavelength of 800 nm (1.55 eV) and a repetition rate of 1 MHz. Pump and probe pulses were focused on the sample at nearly normal incidence with spot size diameter of $\simeq$ 160 and 40 $\mu$m, respectively. Pump and probe fluence were about 25 and 2 $\mu$J/cm$^2$, respectively. Pump and probe beams are $s$- and $p$-polarized, respectively. Data was collected on a freshly cleaved surface from 5 to 300 K. All measurements were carried out under high vacuum (10$^{-6}$ mbar).

Raman spectra were acquired by a Renishaw spectrometer in a back-scattering configuration. A 633 nm laser with spot size of about 1 $\mu$m was the exciting light source. The typical resolution of the system was 1 cm$^{-1}$.

## III. RESULTS AND DISCUSSION

Figure 1 illustrates PdTe$_2$ differential reflectivity $\Delta R/R$ at $T$ = 100 K in linear (main panel) and log (inset) timescales. Electron-electron ($e$-$e$) thermalization and electron-boson scattering processes determine $\Delta R/R$ time evolution. Photoexcitation results in a rapid rise of $\sim$ 0.2 ps. This was followed by two distinct recovery processes. The first, a fast recovery process ($\tau_f$), occurred within 0.3 ps. The second slower relaxation ($\tau_s$) then occurred. After the $\sim$ 40 ps relaxation period, the signal was dominated by a flat offset which is indicated by a dashed blue line. Two significant high-frequency oscillations occurred instantaneously upon photoexcitation and superimposed on the $\Delta R/R$ profile. This simultaneity is evidenced by the small asymmetric periodic wiggles in the Figure 1. Relaxation curves are significantly affected by the large amplitude of the two oscillations. Oscillations and carrier dynamics contributed the first reflectivity peak. The rapid relaxation process $\tau_f$ lifetime is comparable to that of the pumping pulse duration. The signal de-convoluted with a finite width Gaussian pump pulse. The solid red line in Figure 1 suggests data fitting well with

$$\frac{R(t)}{R} = \frac{1}{\sqrt{2\pi}w}\exp(-\frac{t^2}{2w^2}) \otimes [\sum_{i=1,2} A_i\exp(-\frac{t-t_0}{\tau_i}) + \sum_{j=1,2} B_j\exp(-\Gamma_j(t-t_0))\sin(\Omega_j t + \phi_j)] + C \quad (1)$$

where $A_i$ and $\tau_i$ are amplitude, and relaxation times of the $i$th nonoscillatory signal, respectively, and describe carrier dynamics. $B_j$, $\Gamma_j$, $\Omega_j$, and $\phi_j$ are the $j$th oscillatory signal amplitude, damping rate, frequency, and initial phase, respectively. These describe the collective excitation dynamics. $w$ and $C$ are incidence pulse temporal duration and a constant offset, respectively.

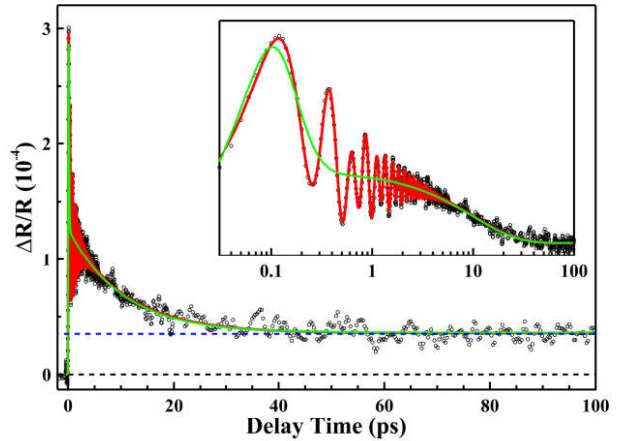

FIG. 1. (Color online)$\Delta R/R$ transient reflectivity is shown as a function of delay time in PdTe$_2$ at 100 K up to 100 ps. Red and green solid lines are the Eq.(1) fits with, and without oscillations, respectively.

Figure 2 focuses on $\Delta R/R$ differential reflectivity at selected temperatures in the 5 - 300 K range over a short time scale. The data are similar at different temperatures with two distinct relaxation processes and two pronounced oscillations. Oscillation amplitudes increased as temperatures dropped. $\Delta R/R$ reflectivity fitted well with the Eq.(1) at all measured temperatures.

First, we focus on the nonoscillatory response, i.e., the two relaxation processes. The quasiparticle dynamic quantitative of the nonoscillatory, or two relaxations, response was studied for temperature-dependent behavior and provide insight into relaxation processes origins. The relaxation processes fitted well to the Eq. (1) when oscillations were not included (set $B_j = 0$). Attracted decay times $\tau_f$ and $\tau_s$ were plotted as a function of temperature (Figure 3). The quick recovery process ($\tau_f < 0.2$ ps) quickened as temperature fell. It saturated at $\sim 100$ K [Fig. 3(a)]. After femtosecond photoexcitation, electron temperatures rose, on a femtosecond time scale, to thousands of K. High-energy hot electrons then transferred excess energy to the lattice at a subpicosecond time scale through carrier-phonon interaction [36, 37]. $\tau_f$ temperature dependence was similar to that of the $e$-$ph$ thermalization observed in other materials [32, 33, 38]. Since $e$-$e$ scattering is much quicker than $e$-$ph$ thermalization, $\tau_f$ is dominated by phonon emissions. It seems reasonable to attribute $\tau_f$ in PdTe$_2$ to electronic system cooling via $e$-$ph$ thermalization rather than to $e$-$e$ scattering.

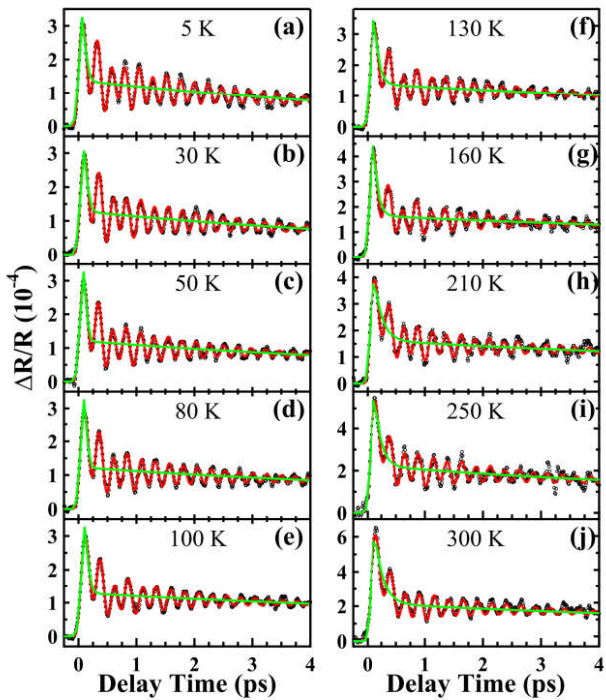

FIG. 2. (Color online) (**a**)-(**j**) $\Delta R/R$ PdTe$_2$ transient reflectivity at several selected temperatures between 5 and 300 K. The solid red and green lines fit measured signal, with and without the oscillations, respectively.

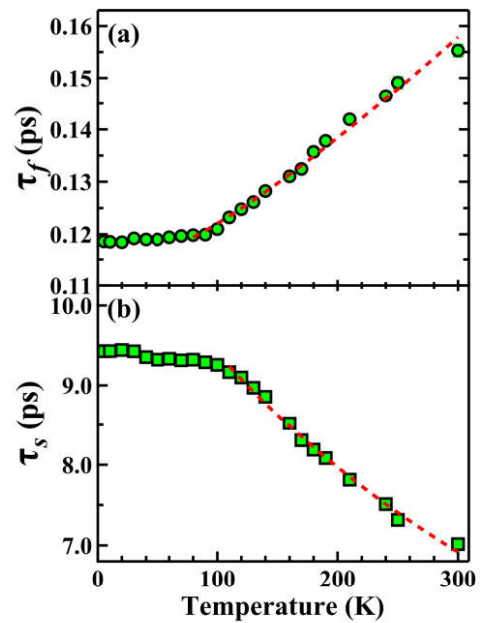

FIG. 3. (Color online) (**a**) shows $\tau_f$ while (**b**) shows $\tau_s$, as a function of temperature. The red dashed curves denote (a) the TTM fit and (b) the fit using Eq.(4).

A two-temperature model (TTM) is usually used to describe $e$-$ph$ thermalization temperature dependence [32, 33, 38]. At low temperatures, $e$-$ph$ and $e$-$e$ have comparable thermalization times. This unsuits TTM as a model for studying low temperature $e$-$ph$ thermalization. PdTe$_2$ $\Theta_D$ is about 210 K [39]. According to the TTM mode noted in the literature [32, 33, 38], temperature dependent data above 80 K was fitted [41]. As shown by the dashed red line in Figure 3(a), a very good fitting result is obtained. This results supports attributing $\tau_f$ to $e$-$ph$ thermalization [41]. Taking optical constants with optical conductivity $\sigma_1 \sim 2.63 \times 10^3$ $\Omega^{-1}$cm$^{-1}$ and reflectivity $R = 0.75$ with 800 nm photons at 300 K by Heumen *et al.* [40], penetration depths and laser energy densities were determined to be $l_s \simeq 25$ nm [42] and $U_l \simeq 2.23$ J/cm$^3$. Cryostat window transmission is taken into account. The electronic specific heat coefficient and $e$-$ph$ coupling constant can be obtained from TTM fit: $\gamma \sim 9.5$ Jm$^{-3}$K$^{-2}$ and $g_\infty \sim 3.2 \times 10^{16}$ Wm$^{-3}$K$^{-1}$.

According to the Allen's electron thermal relaxation model [36], relaxation time, $\tau$, of the quasiparticle in metals is given by:

$$\frac{1}{\tau} = \frac{3\hbar\lambda\langle\omega^2\rangle}{\pi k_B T_e} \qquad (2)$$

where $\lambda$ is the $e$-$ph$ coupling constant, $\lambda\langle\omega^2\rangle$ is the second momentum of Eliashberg function. $T_e$ can be further described by [43, 44]

$$T_e = \langle\sqrt{T_l^2 + \frac{2(1-R)F}{l_s\gamma}e^{-z/l_s}}\rangle \qquad (3)$$

where $R$ is reflectivity at 800 nm; $F$ is pumping fluence; $\gamma$ is the electronic specific heat capacity coefficient; and, $z$ is the depth away from the sample surface. At 300 K, $T_e \simeq 619$ K is obtained. From the measured relaxation time at 300 K, $\tau_f \sim 0.155$ ps, $\lambda\langle\omega^2\rangle \sim 5.47 \times 10^{26}$ Hz$^2$ was derived. Taking into account the $E_g$ ($\sim 2.6$ THz at 5 K) and $A_{1g}$ ($\sim 4.2$ THz at 5 K) modes obtained below, namely $O_2$ and $O_3$ modes proposed by theoretical calculations [30], the nominal $e$-$ph$ coupling constant are $\lambda_{F_g}$ $= 2.05$ and $\lambda_{A_{1g}} = 0.79$, respectively. $\lambda_{E_g}$ is significantly larger than the results of recent theoretical (0.57) [30], electronic transport (0.59) [45], and helium atom scattering (0.58) [46] studies. It may be because $<\omega_{E_g}^2>$ is much smaller than PdTe$_2$ $<\omega^2>$. $\lambda_{A_{1g}}$ is slightly larger than recent results [30, 45, 46], indicating that the $A_{1g}$ mode may provide a large contribution to superconductivity, which is also consistent with the theoretical calculation [30]. A slightly larger value of $\lambda_{A_{1g}}$ also suggests that to understand the BCS-type superconductivity in PdTe$_2$, it is necessary to consider the involvement of higher frequency phonons in the formation of Cooper pairs.

Figure 3(b) illustrates slow component $\tau_s$ temperature dependence. Unlike $\tau_f$, $\tau_s$ increased as temperature dropped until saturating at $\sim 80$ K. After the initial $e$-$ph$ thermalization, excess electrons (holes) in the conduction (valence) bands would have moved closer to the Fermi energy ($E_F$). The origin of $\tau_s$ is discovered by examining Fermi Surface (FS) topologies and band structure of PdTe$_2$. Earlier ARPES measurements indicated that PdTe$_2$, like other semimetals [32, 47], coexists with electron and hole Fermi pockets [16–21]. Naturally, $\tau_s$ might originate from an $e$-$h$ recombination between conduction and valence bands. This process exists widely in semimetals. $\tau_s$ relaxation times are $7 \sim 9.5$ ps and much faster than that of $e$-$h$ radiative recombination which is a few nanoseconds [48]. Electron and hole Fermi pockets are separated in momentum space. Momentum conservation requires that $e$-$h$ recombination processes be assisted by a third quasiparticle. It seems most likely that a phonon-assisted $e$-$h$ recombination between electron and hole pockets is involved.

According to prior theoretical and experimental works on band structures [16, 20], varieties of possible electron transitions can be induced by 800 nm laser, including bulk and surface state electrons. For example, along the $\Gamma$-$A$ direction, some electrons can be excited directly from the valence band to the unoccupied state near the Fermi level. Phonon-assisted $e$-$h$ recombination has been observed in other semimetals with similar temperature dependence behavior [32, 35]. It has been proposed that the $e$-$h$ recombination time can be quantitatively described by [32, 49]

$$\frac{1}{\tau} = A\frac{\frac{\hbar\omega}{2kT}}{sinh^2(\frac{\hbar\omega}{2kT})} + \frac{1}{\tau_0} \qquad (4)$$

where $\omega$ is the frequency of the phonon mode assisting $e$-$h$ recombination, $\tau_0$ is $T$-independent recombination time, and $A$ is the fitting parameter. The red dashed line in Figure 3(b) suggested that $\tau_s$ fits well with Eq.(4) at high temperatures. This is consistent with the theoretical expectation that lower temperature leads to a reduction in phonon population, which reduces the efficiency of interband $e$-$h$ scattering and leads to a longer $e$-$h$ recombination time.

Next, PdTe$_2$ collective excitations dynamics, superimposed on $\Delta R/R$ curves (Figure 1 and Figure 2), were analyzed. Optical phonon modes can be extracted by directly fitting a total dynamics trace with the Eq. (1) (Figure 1 and Figure 2, solid red lines), or analyzing the curves after subtracting the quasiparticle dynamics (Figure 1 and Figure 2, solid green lines) from the total-dynamics trace. This study used the latter approach. The time-domain oscillations for several selected temperatures appear in Figure 4(a). The red curves are the fitting curves based on a two-components damped oscillation function. As usual, fast Fourier transform (FFT) was used to extract the coherent collective excitations in the frequency domain. Two distinct high frequency terahertz (THz) modes, $\Omega_1$ and $\Omega_2$, were resolved at all measured temperatures [Fig. 4(b)]. These two phonons are consistent with previous phonon spectra calculation, $\Gamma$-center optical mode $O_2$ and $O_3$ [30]. A lower frequency $\Omega_1$ mode, frequency domain data is relatively dispersed possibly due to intrinsically low phonon signal, but also because its cycle is about twice that of $\Omega_2$. During the carriers relaxation processes, oscillations caused by collective excited states such as coherent phonons and charge density waves in the material are further manifested as reflectivity signal oscillations [50–52]. Usually, oscillatory components persisting at room temperatures are attributed to coherent phonons. The phonons are initiated via dispersive excitations [53] or photoinduced Raman modes.

Raman scattering results are shown in Figure 4(c). At least two significant vibrational modes were resolved. At room temperature (300 K), the two Raman shift modes values are $\sim 75.4$ cm$^{-1}$ and $\sim 133.2$ cm$^{-1}$. These correspond to Te atoms in-plane ($E_g$) and out-of-plane ($A_{1g}$) motions [54, 55], respectively. These values are consistent with previous reports bulk PdTe$_2$ values [55]. Figure 4(b) and (c) are plotted on the same horizontal axis range. $\Omega_1$ and $\Omega_2$ Frequencies are identical to that of $E_g$ and $A_{1g}$ phonons, respectively. This suggests that $\Omega_1$ and $\Omega_2$ originate from the $E_g$ and $A_{1g}$ phonons, respectively. The intensities of both the ultrafast optical and Raman scattering measurements depend on the configuration, including polarization and angle, of the incident and scattered photons. The significant differences in the relative phonon intensities in the time domain and frequency domain may be due to the different configuration of the two measurements.

Figure 4(d) and (e) display phonon frequencies as a function of temperature. $\Omega_1$ and $\Omega_2$ frequencies were derived from peak positions of FFT spectra. $E_g$ and $A_{1g}$ frequencies were estimated by fitting the Raman spectra peak with a Gaussian lineshape. At 5 K, $\Omega_1$ and $\Omega_2$ frequencies were about 2.6 and 4.2 THz, respectively. $\Omega_2$ ($A_{1g}$) mode frequency obtained from both ultrafast spectroscopy and Raman measurements revealed that it decreased steadily as temperatures increased. When tem-perature increased from 5 to 300 K, $\Omega_2$ mode redshifted about 0.2 THz. Instead of a monotonic decrease as temperature rose, $\Omega_1$ had essentially stabilized at temperatures above 80 K. $E_g$ measured by Raman scattering changes very little at high temperatures. The damping rates were extracted by fitting the oscillations. The temperature dependence of damping rates $\Gamma_1$ and $\Gamma_2$ of $\Omega_1$ and $\Omega_2$, respectively, are plotted in Figure 4(f). $\Gamma_2$ damping rate changed little over the measured temperature range. $\Gamma_1$ decreased as temperature increased.

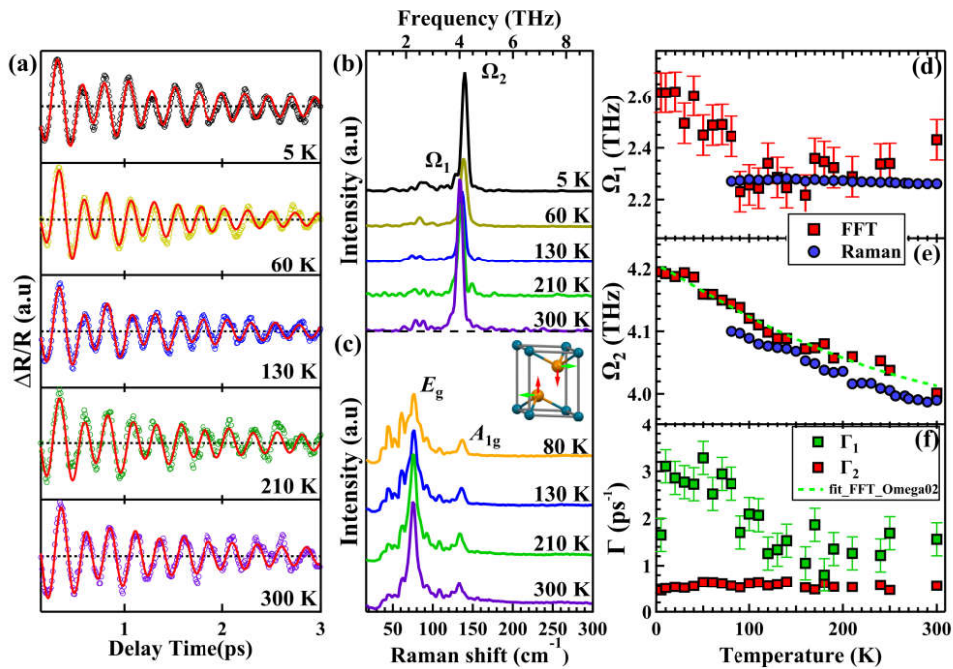

FIG. 4. (**a**) Oscillations after subtracting nonoscillatory background for five temperatures. Fitted results are in red. (**b**) FFT frequency-domain data corresponding to (a). Curves are shifted vertically for discernibility. (**c**) PdTe$_2$ Raman spectra at various temperatures. Inset: Schematic representation of the motion of Te atoms corresponding to in-plane $E_g$ (green arrows) and out-of-plane $A_{1g}$ (red arrows) modes. (**d**) and (**e**) denotes $\Omega_1$ ($E_g$) and $\Omega_2$ ($A_{1g}$) temperature dependence, respectively. The frequencies obtained from FFT and Raman shift are represented by red squares and blue circles, respectively. The green dashed line is the fitted results using an anharmonic phonon model. (**f**) The derived damping rate $\Gamma_1$ and $\Gamma_2$ of coherent optical phonon $\Omega_1$ and $\Omega_2$, respectively.

In general, population decay occurs along with pure dephasing determined coherent phonon damping. Previous studies suggested that coherent phonon decay processes in semimetal and topological insulators are largely determined by population decay caused by anharmonic phonon-phonon coupling [33, 35, 56, 57]. Phonon population is a strong function of temperature. Coherent phonon frequency and damping rates are strongly temperature-dependent. For anharmonic effects, usually, the frequency decreases as temperature rises while damping rate increases [33]. As shown by the green dashed line in Figure 4(e), optical phonon anharmonic effects can explain the temperature dependence of $\Omega_2$. However, this is not the case with modes $\Omega_1$, which cannot be well explained in terms of anharmonic effects alone.

Pure dephasing always exists because a quantum system is always connected to its surroundings. Previous studies have shown that PdTe$_2$ is a BCS-type superconductor. Pure dephasing, e.g., *e-ph* scattering, should accounted for.

## IV.CONCLUSIONS

In this study, ultrafast time-resolved differential reflectivity and Raman scattering measurements were performed on PdTe$_2$ as a function of temperature to reveal important carriers and coherent phonons roles. Two exponential relaxation processes with distinctive time constants were readily observed. The origins of these two processes can be attributed to *e-ph* thermalization ($\tau_f$ < 0.2 ps) and phonon-assisted *e-h* recombination ($\tau_s \sim$ 7-9.5 ps). Two vibrational modes were resolved as originating from Te atoms in-plane ($E_g$) and out-of-plane ($A_{1g}$) motions. Both $E_g$ and $A_{1g}$ phonon mode softens, redshift, as temperatures rise from $\sim$ 2.6 and 4.2 THz at 5 K to $\sim$ 2.3 and 4.0 THz at 300 K, respectively. The *e-ph* scattering plays important roles in the relaxation process. Our analysis suggests that the $A_{1g}$ mode contributes greatly to superconductivity, and higher frequency phonons are also involved in the formation of Cooper pairs, which must be considered in understanding the superconductivity of PdTe₂.

## ACKNOWLEDGMENTS

We are grateful for valuable discussions with J. Qi, C. C Shu, and M. X Chen. Our work was supported by the National Natural Science Foundation of China (Grants No. 12074436, and No. U2032204), and Chinese National Key Research and Development Program (2016YFA0300604). J. Q Meng would like to acknowledge support from the Innovation-driven Plan in Central South University (2016CXS032).

---